\documentclass[preprint,journal]{vgtc}       

\ifpdf
  \pdfoutput=1\relax                   
  \usepackage{graphicx}                
  \DeclareGraphicsExtensions{.pdf,.png,.jpg,.jpeg} 
\else
  \usepackage{graphicx}                
  \DeclareGraphicsExtensions{.pdf}     
\fi%

\graphicspath{{figures/}{pictures/}{images/}{./}} 

\usepackage{microtype}                 
\PassOptionsToPackage{warn}{textcomp}  
\usepackage{textcomp}                  
\usepackage{mathptmx}                  
\usepackage{times}                     
\usepackage{cite}

\usepackage{amsmath}
\usepackage{amssymb}
\usepackage{mathrsfs}

\usepackage{color}
\usepackage{subfig}
\usepackage{url}
\usepackage{wasysym}
\usepackage[pagebackref=true]{hyperref}

\newcommand{\red}[1]{}

\newcommand{\eg}[1]{{\emph{e.g.}, }}

\onlineid{270}

\vgtccategory{Research}

\title{Improved characterization of Lagrangian coherent structures through time-scale analysis}

\author{Zi'ang Ding and Xavier Tricoche}
\authorfooter{
\item
 Zi'ang Ding is with Microsoft. E-mail: ziang.ding@microsoft.com.
\item
 Xavier Tricoche is with Purdue University. E-mail: xmt@purdue.edu.
}

\shortauthortitle{Ding \MakeLowercase{\textit{et al.}}: Improved characterization of Lagrangian coherent structures through time-scale analysis}

\abstract{The computation of Lagrangian coherent structures (LCS) has established itself as a prominent means to reveal significant geometric structures in time-dependent vector fields. Their characterization, however, requires the selection of a suitable time parameter for the construction of the flow map that may not be known in advance. We present in this paper a continuous time-scale framework for LCS extraction and visualization. Specifically, we treat the time axis as a continuum from which a best temporal scale is automatically determined at each spatial location for the extraction of LCS. Beyond its effectiveness with vector fields we show that this method can be successfully applied to improve the characterization of salient structures in tensor fields and discrete maps. We present applications of our method to problems spanning fluid dynamics, medical imaging, and orbital mechanics. The results show that our approach can reveal important structural features that are missed by existing LCS extraction methods.
} 


\CCScatlist{ 
 \CCScat{K.6.1}{Management of Computing and Information Systems}%
{Project and People Management}{Life Cycle};
 \CCScat{K.7.m}{The Computing Profession}{Miscellaneous}{Ethics}
}

   \teaser{
   \centering
   \includegraphics[width=16cm]{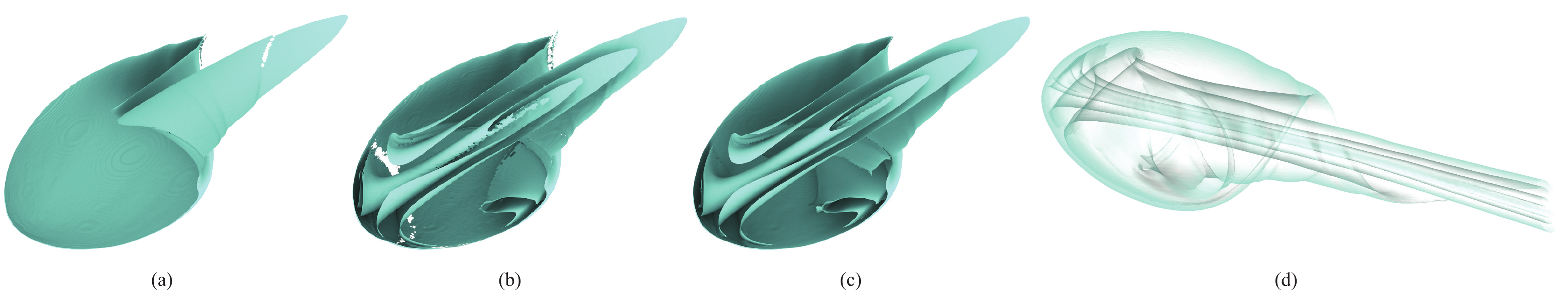}
   \caption{A comparison of fixed integration time approach and our approach on the right flow recirculation bubble of the \textit{Delta Wing} dataset. (a) Ridge surfaces extracted from a FTLE field computed with a ``short'' integration time $\tau=0.010$; (b) ridge surfaces corresponding to a ``long'' integration time $\tau=0.025$; (c) ridge surfaces extracted by our approach over an integration time interval $[0.005, 0.025]$; (d) ridge surfaces corresponding to (c) rendered with transparency. Ridge surfaces extracted from a ``short'' integration time miss inner parts of the flow recirculation bubble, which appear after a ``long'' integration time, while a ``long'' integration time yields a low-quality approximation of the geometry of the outer layer of the recirculation bubble. In contrast, our proposed temporal-scale approach correctly reconstructs the LCS of both inner and outer parts.}
   \label{fig:delta_wing_right_result}
  }

\vgtcinsertpkg

\begin{document}


\firstsection{Introduction}

\maketitle

Lagrangian coherent structures (LCS) correspond to material surfaces embedded in time-dependent flows that form the boundaries of regions exhibiting distinct dynamics behavior. As such, they offer an effective means to understand the geometric structure and visualize the transport properties of complex fluid flows. A popular characterization of LCS is based on the \textit{finite-time Lyapunov exponent} (FTLE), which measures the local rate of separation of nearby fluid particles: high values of FTLE for forward advection indicate a repelling behavior while high values during backward advection signal an attracting behavior. Practically a repelling LCS is then extracted as a ridge of the forward-time FTLE field and an attracting LCS as a ridge of the backward-time FTLE field~\cite{Haller2000b, Haller2001}. Together, these structures control the behavior of the flow and nearby massless tracers.

\red{
We need to convey two things in this paragraph:
1. finding the proper integration time for FTLE / LCS computation is non-trivial and costly
2. In many practical cases, there is no single time scale that can correctly capture all the structures of interest. One therefore needs a spatially varying integration time.}

The FTLE-based approach to characterizing LCS has proved successful in a wide array of fluid dynamics applications~\cite{Peacock2010}. Each time, a suitable advection time must be specified by the user before computing the FTLE measure through numerical integration of the vector field. Yet, in many instances, a good integration time may not be known a priori, which leads to costly trials before a satisfactory parameter value can be determined. Indeed, the integration time should be chosen long enough to allow flow features to emerge. On the other hand, it should not be too long to prevent multiple flow structures to compound their individual signatures in the FTLE field and create aliasing artifacts as a result. In addition, since most vector fields in computational fluid dynamics are simulated within a bounded domain, the motion of particles is undefined once they leave the domain, which complicates the FTLE computation. In many applications, the integration time used to compute FTLE must therefore be carefully chosen based on prior knowledge of the flow behavior or on multiple and possibly redundant computations.

\red{we need to acknowledge that there is prior work on the topic and explain why we are adding another paper to the list. What problem(s) with these methods are we addressing?}

Considering the time axis as a continuum, we introduce a temporal-scale analysis approach to determine the best integration time at each spatial location, and thereby extract ridges over different integration times. The main contributions of this paper include:
\begin{itemize}
  \item Automatic detection of the spatially varying best integration time to characterize all relevant LCS. 
  \item Explicit characterization and combined visualization of long-time and short-time FTLE ridges.
  \item Improved structure characterization over existing LCS methods.
  \item Extension of the time-scale concept to the characterization of topological structures in chaotic maps.
\end{itemize}

The paper is organized as follows. Section~\ref{sec:2} reviews past work that is most directly related to our approach. We document the aliasing issues associated with the extraction of LCS from FTLE fields computed with uniform integration length with a well-known example in section~\ref{sec:3}, 
before describing our temporal-scale analysis in section~\ref{sec:4}. We describe an extension of the time-scale concept to the study of \textit{maps} in Hamiltonian system in section~\ref{sec:5}. We present results for a 
variety of applications and offer quantitative comparisons with existing methods in section~\ref{sec:6}.
We comment on our findings in section~\ref{sec:7}. Specifically, we provide data about the performance of the proposed 
temporal-scale approach (section~\ref{sec:performance}), the influence of the temporal resolution on the 
results (section~\ref{sec:temporal_resolution}), and the integration of our approach into existing LCS based flow visualization methods (section~\ref{sec:integration}).
Finally conclusion and future research directions are discussed in section~\ref{sec:8}.

\section{Related Work}
\label{sec:2}
Haller pioneered the use of the finite-time Lyapunov exponent (FTLE) to characterize Lagrangian coherent structures in planar and three-dimensional unsteady flows~\cite{Haller2000a,Haller2000b,Haller2001}. A study of the robustness of the LCS characterized by FTLE was presented by the same author~\cite{Haller2002}. Shadden \emph{et al.} showed that LCS can be approximated by maximum ridges of FTLE fields with a sufficiently long integration time as the net flux across those ridges is small and typically negligible~\cite{Shadden2005}. Later, Sadlo et al. presented a method to determine a uniform lower bound of finite time scope for FTLE computation~\cite{sadlo2012}. For a finite domain, Tang \emph{et al.} proposed a method to solve the problem of trajectories leaving the domain by extending the given velocity field~\cite{tang2010}. Haller presented a study showing shortcomings of the FTLE-based approach in the presence of shear and proposed a variational characterization of LCS~\cite{Haller2011}. 
A FTLE benchmark method measuring the quality of the extracted LCS by comparison with a ``ground truth'' was proposed by Kuhn \emph{et al.}~\cite{Kuhn2012}.

\red{the second paragraph should discuss prior work on the issue of time scale: FSLE, Lyapunov time, max FTLE, ...}

Developed as a diagnostic for multi-scale mixing, the \emph{finite-size lyapunov exponent} (FSLE) has also been used as an alternative to FTLE to detect coherent structures in dynamical systems~\cite{aurell1997, boffetta2001}. The basic idea of this method consists in measuring the time needed to achieve a given separation from each location, thereby substituting a \emph{size} parameter to the time parameter found in the definition of FTLE. Direct comparisons between FTLE and FSLE have been made by Boffetta \emph{et al.}~\cite{boffetta2001}, Sadlo and Peikert~\cite{Sadlo2007}, and Peikert \emph{et al.}~\cite{peikert2014}. Most recently, Karrasch and Haller showed that FSLE was prone to inaccuracies and false positives in the characterization of LCS~\cite{Karrasch:2013:Do-finite-size}.

Recently, instead of using a fixed time, a varying integration time concept named Lyapunov time 
was introduced by Sadlo~\cite{sadlo2015} in trajectory-based visualization.

\red{the following 3 paragraphs are mostly off topic... while there is a conceptual connection between our work and prior work on scale space in image processing, we are not dealing with that aspect here. In addition, a paper by one of Peikert's students mentioned that spatial scale seems to have only a marginal effect on the characterization of LCS. That goes back to the idea that LCS need to be sharp to act as material boundaries.}
In image processing and computer vision, the notion of scale-space embeds a signal $f(x)$ into a family of functions $F(x,t)$ with a continuum scales $t$ for analysis. The basic insight behind this approach is that objects inherently comprise details at various scales. Hence the features associated with different observation scales can be detected by extending the original image to a family of smoothed images where the size of the smoothing kernel is defined by the scale~\cite{Lindeberg1994,Lindeberg1996}. A number of feature-extraction methods have been developed within the theoretical framework of scale-space. Lindeberg in particular introduced a ridge detection approach that automatically determines the optimal scale based on a normalized ridge strength criterion at each point~\cite{Lindeberg1998}.

In the visualization community, scale-space theory has been applied in flow visualization and feature extraction. Bauer and Peikert proposed to use \emph{parallel vector operator}~\cite{Peikert1999} to recover vortex lines in scale-space~\cite{Bauer2002}. Klein and Ertl presented a critical points tracking method for vector fields through discretely sampled scales to filter out noise in the dataset~\cite{Klein2007}. Combining scale-space concept with particle systems, Kindlmann \emph{et al.}~\cite{Kindlmann2009} extracted crease surfaces in three-dimensional medical images while maintaining the spatial continuity of the scale.

Recently, two works applying scale-space theory to extract ridges in FTLE fields have been published. Barakat and Tricoche~\cite{Barakat2010} proposed a GPU-based adaptive technique for height ridge~\cite{eberly1996} extraction and visualization across scales, while Fuchs \emph{et al.}~\cite{Fuchs2012} introduced a similar method to characterize the C-ridge~\cite{Schindler2012}. Though an optimal integration time concept was mentioned, Fuchs \emph{et al.} did not explore this avenue further in their work.


\section{Background}
\label{sec:3}
In this section, we review the widely used FTLE-based LCS definition proposed by Haller~\cite{Haller2000a}. Further, we illustrate the possible issues associated with a uniform integration time in structure characterization by considering a well known example.

\subsection {Finite-Time Lyapunov Exponent and LCS}

The trajectory seeded at $(\mathbf{x_{0}},t_{0})$ is the solution $\phi_{\mathbf{x_{0}},t_{0}}(t)$ of following initial value problem in a given time-dependent vector field $\mathbf{u}(\mathbf{x}, t)$

\begin{equation}
\label{eq:ode}
\left\{\begin{array}{rcl}
\frac{\partial }{\partial t}\phi_{\mathbf{x_{0}},t_{0}}(t) & = & \mathbf{u}(\phi_{\mathbf{x_{0}},t_{0}}(t),t) \\ \phi_{\mathbf{x_{0}},t_{0}}(t_{0}) & = & \mathbf{x_{0}},
\end{array}\right.
\end{equation}

whereby $t=t_{0}+\tau$ and $\tau$ is the integration time. Keeping $t_{0}$ and $\tau$ fixed and varying $\mathbf{x_{0}}$, one obtains the flow map $\phi_{t_{0}}^{t}(\mathbf{x})$. The spatial variations of this flow map around $\mathbf{x_{0}}$ are determined by its Jacobian $J_{\mathbf{x}}(t,t_{0},\mathbf{x_{0}}):=\nabla_{\mathbf{x_{0}}} \phi_{t_{0}}^{t}(\mathbf{x_{0}})$ and the maximal dispersion around $\mathbf{x_{0}}$ at $t$ is given by the spectral norm of $J_{\mathbf{x}}(t,t_{0},\mathbf{x_{0}})$:

\begin{equation}
\sigma_{\tau}(t_{0},\mathbf{x_{0}}):=\sqrt{\lambda_{max}(J_{\mathbf{x}}(t,t_{0},\mathbf{x_{0}})^{T}J_{\mathbf{x}}(t,t_{0},\mathbf{x_{0}}){})}.
\end{equation}

Normalizing by the integration time $\tau$ yields the \emph{finite-time Lyapunov exponent}:

\begin{equation}
\mathbf{FTLE}(t_{0},\mathbf{x_{0}},\tau)=\frac{1}{|\tau |}ln\sigma_{\tau}(t_{0},\mathbf{x_{0}}).
\label{eq:ftle}
\end{equation}

Lagrangian coherent structures can then be extracted as ridges of the FTLE field, whereby repelling LCS are characterized by forward FTLE and attracting LCS are revealed by backward FTLE.

\subsection {Limitations of Fixed Time FTLE for LCS Extraction}

\begin{figure}[!h]
   \centering
   \includegraphics[width=\linewidth]{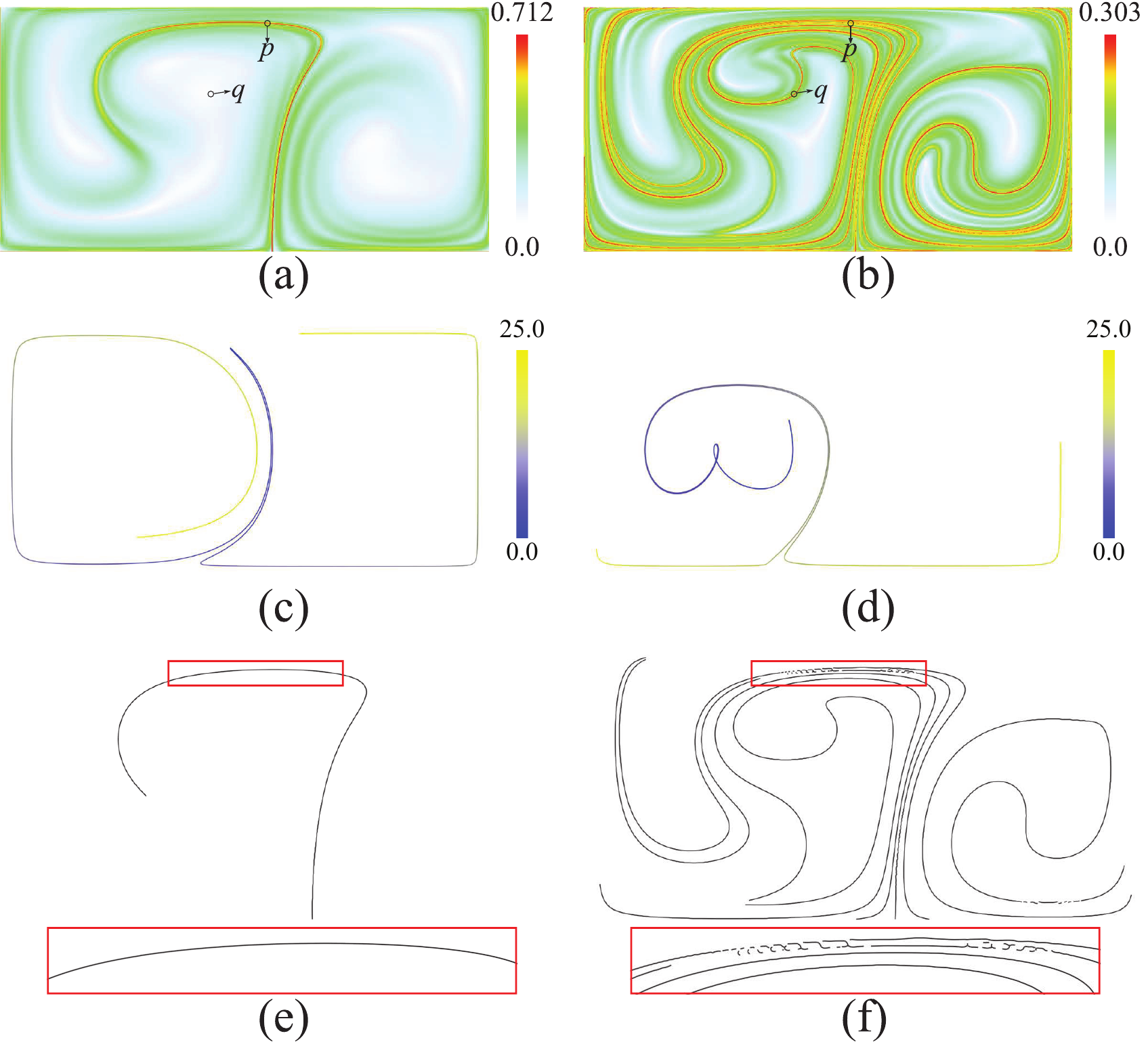}
   \caption{(a) and (b) Forward FTLE fields computed at two different integration times with $\tau=9$ and $\tau=22$ from the ``\textit{Double Gyre}" vector field; (c) and (d) trajectories of particles near sampling point $p$ and $q$ marked in image (a) and (b). (e) and (f) extracted ridge lines from image (a) and (b). Trajectories are color coded by the integration time ranging from $0.0$ (blue) to $25.0$ (yellow).\label{fig:standard_ftle_double_gyre}}
 \end{figure}

Images (a) and (b) in Figure~\ref{fig:standard_ftle_double_gyre} show the variation of forward FTLE fields with two different integration time $\tau=9$ and $\tau=22$ from the \emph{Double Gyre} vector field~\cite{Shadden2005}. Comparing both images, one may notice that as the integration time $\tau$ is increased, several ridges are revealed by the FTLE field while others become obfuscated by aliasing. In this paper, we define the short-time and long-time FTLE ridge as the one associated with an FTLE field computed with a short integration time and long integration time respectively. Two sampling points $p$ and $q$ on the short-time and long-time ridges were selected respectively. Image (c) and (d) show trajectories of particles released near these sampling points. Particles released near sampling point $p$ stay close to each other until they have reached the saddle point at the bottom and separate from each other at an integration time close to $7$. These particles start to come close to each other again as they are effected by the saddle point at the top at an integration time close to $19$ and cause the aliasing effects on the FTLE field. In contrast, particles released near sampling point $q$ do not reach the saddle point at the bottom until the integration time approaches $20$. Therefore, a single fixed integration time fails to characterize LCS at both sampling point $p$ and $q$. Image (e) and (f) show the extracted LCS as ridge lines from each FTLE field. Though the result for $\tau=22$ contains short-time LCS as well as long-time LCS, the characterized short-time LCS are broken and discontinuous compared to the one obtained for $\tau=9$.

In general, LCS can shrink, grow, appear, and disappear with changes in integration time~\cite{Shadden2011}. Therefore, a single and fixed integration time is not sufficient to simultaneously characterize and visualize LCS that develop over different time scales in flow visualization applications. Different integration time must be specified for short-time structures and long-time structures. In this work, we consider the integration time as a degree of freedom within a time interval defined by the user, and let the system automatically determine the optimal value. The following section presents our temporal-scale, which follows this basic principle.

\section{Temporal-Scale approach}
\label{sec:4}
 In this section we describe an approach that characterizes LCS with different life spans
through a temporal-scale analysis. With a specified starting time $t_{0}$ and a user
defined integration time interval $[\tau_{LB}, \tau_{UB}]$, a series of FTLE fields
$\mathscr{F}_{t_{0}}^{\tau_{i}}$ could be computed for various values of the integration time
$\tau_{i} \in [\tau_{LB}, \tau_{UB}]$. Our temporal-scale approach analyzes these
FTLE fields and extracts ridges from them.

In digital image processing, edges can be detected through a scale analysis method~\cite{DIPA}.
First a feature map which records the maximum edge strength of each pixel from different
scales is created through the scale analysis.
Then edges are characterized by extracting ridges from this feature map.
Inspired by it, our approach consists of the two following steps,

\begin{enumerate}
  \item generates a hyper-FTLE field through a temporal-scale analysis on all FTLE fields $\mathscr{F}_{t_{0}}^{\tau_{i}}$ computed within the user defined interval $[\tau_{LB}, \tau_{UB}]$;
  \item characterizes LCS directly from the hyper-FTLE field.
\end{enumerate}

\subsection {Hyper-FTLE field generation}
\label{sec:hyper_ftle}
The FTLE-based LCS definition~\cite{Haller2001} states that in forward time a large FTLE
value indicates strong divergence while in backward time a large FTLE value designates strong
attraction. Therefore, a ridge associated with a larger FTLE value is presumably more significant
than one associated with a smaller FTLE value.
Furthermore, a study by Shadden et al.~\cite{Shadden2005} shows that sharp, well-defined
ridges are more Lagrangian than poorly defined ridges, whereby the sharpness is measured by the ridge strength.
Based on the above two points, here we define our hyper-FTLE field as

\begin{equation}
\mathscr{H}_{t_{0}}(\mathbf{x}) = \max_{\tau_{i} \in [\tau_{LB}, \tau_{UB}]}(\mathscr{F}_{t_{0}}^{\tau_{i}}(\mathbf{x})\times \mathscr{S}_{t_{0}}^{\tau_{i}}(\mathbf{x}))
\end{equation}

where $\mathscr{S}_{t_{0}}^{\tau_{i}}$ is the ridge strength estimated from the FTLE field
$\mathscr{F}_{t_{0}}^{\tau_{i}}$. The integration time $\tau_{i}$ which maximizes
$\mathscr{F}_{t_{0}}^{\tau_{i}}(\mathbf{x})\times \mathscr{S}_{t_{0}}^{\tau_{i}}(\mathbf{x})$
is defined as the \emph{optimal integration time} in this paper. Although there are many possible ways
to combine FTLE value with ridge strength into a quantity to maximize,
the simple solution here not only produces highly accurate LCS (section~\ref{sec:double_gyre_result}),
but also prevents problematic spatial discontinuities that would interfere with the
ridges that we are interested in.

In this paper, without loss of generality, we consider the height ridge as our ridge definition.
According to this definition, a point lies on the ridge where the field reaches a maximum in a transversal direction corresponding to the minor eigenvector of the
\textit{Hessian matrix}, and the minor eigenvalue is negative. The ridge strength
is quantified by the magnitude of the minor eigenvalue of the
\textit{Hessian matrix}~\cite{Lindeberg1998}.

In contrast to previous techniques~\cite{Kindlmann2009, Barakat2010, Fuchs2012}, our approach reduces the
memory requirements by only recording the hyper-FTLE value and the optimal integration time at each point during the
computation.

For every FTLE field $\mathscr{F}_{t_{0}}^{\tau_{i}}$ computed with integration time $\tau_{i}$,
\textit{Hessian matrix} $H$ is evaluated by central differences, followed by the ridge strength computation.
Then for each point on the sampling grid, the temporal-scale analysis approach updates the maximal hyper-FTLE
value as well as the corresponding integration time if necessary.
Figure~\ref{fig:double_gyre_result}(a)  shows the forward time hyper-FTLE
field of \textit{Double Gyre} dataset with a user defined integration time interval $[1, 25]$, and
figure~\ref{fig:double_gyre_result}(b) visualizes the corresponding optimal integration time determined
by the temporal-scale analysis.

\subsection {Ridge extraction}
\begin{figure}[!h]
  \centering
  \includegraphics[width=\linewidth]{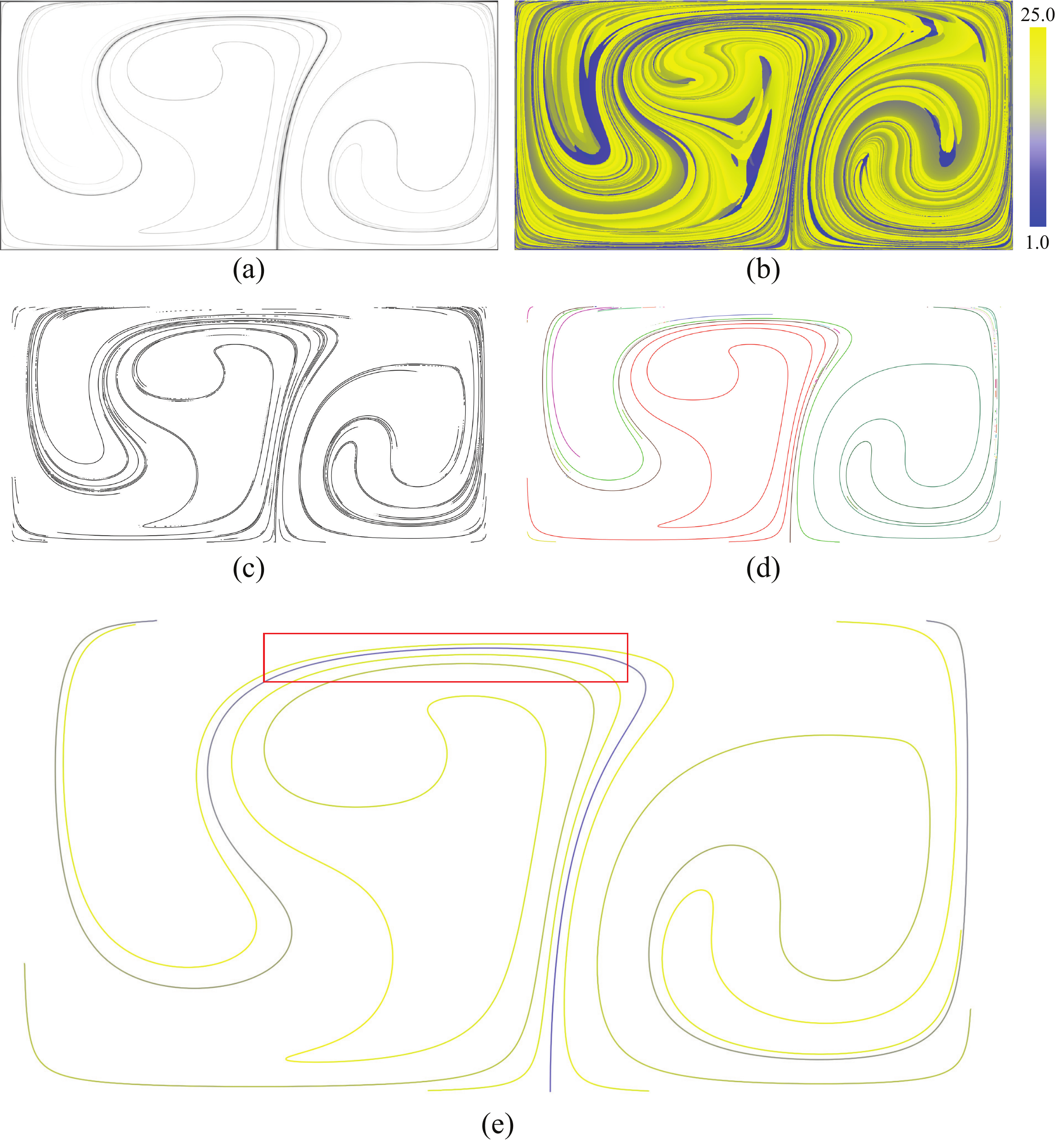}
  \caption{The ridge extraction result on \textit{Double Gyre} dataset with a user specified integration time interval $[1, 25]$. (a) Hyper-FTLE field created by our approach; (b) optimal integration time determined by our approach; (c) ridge line segments computed from (a); (d) connected ridge lines, color coded by a random color; (e) ridge lines color coded by the optimal integration time, ranging from $1(blue)$ to $25(yellow)$.\label{fig:double_gyre_result}}
\end{figure}

After the hyper-FTLE field is generated, one can apply the original nonparallel Marching Ridges method~\cite{Jacob2001}
to extract ridges from it. In order to accelerate the ridge extraction process, here we present a modified Marching
Ridges method consisting of the following three steps.
\begin{enumerate}
  \item For every cells in the sampling grid, determines average directions, calculates zero-crossings, and checks the minor eigenvalue $\lambda_{min}$ in parallel. This step collects all ridge segments (line segments in 2D, triangles in 3D) for further computation. Similar to the hyper-FTLE generation, 1st and 2nd order derivatives are estimated though central differences. Figure~\ref{fig:double_gyre_result}(c) shows the ridge line segments computed from Figure~\ref{fig:double_gyre_result}(a).
  \item Connects all ridge segments after filtering with a threshold on ridge strength. Figure~\ref{fig:double_gyre_result}(d) colors each connected ridge line by a unique random assigned color.
  \item The final ridges are extracted by filtering with a threshold on length or area. Figure~\ref{fig:double_gyre_result}(e) shows the final ridge lines extracted by our approach. The ridge lines are color coded by the optimal integration time determined through the temporal-scale analysis. The smooth transition of colors along the extracted ridge lines shows the smoothness of the time-selection along them.
\end{enumerate}

\section{Temporal-Scale analysis in maps}
\label{sec:5}
\textit{maps} can be thought of as describing successive states of a discrete dynamical 
system. From a visualization standpoint, maps are
hard to study because of their fractal topological structures and regions
of chaotic behavior~\cite{Tricoche2011}.

Recently, the definition of FTLE was extended to such maps in the context of astrodynamics problems to investigate the underlying
salient structure of the dynamics~\cite{gawlik2009lagrangian}. Due to the
discrete nature of the dynamics in this case, the notion of finite-time
must be expressed in number of iterations and the \emph{finite iteration Lyapunov exponent} (FILE) at a given location $\mathbf{x_{0}}$ is defined as:

\begin{equation}
\mathbf{FILE}(0,\mathbf{x_{0}},k)=\frac{1}{|k|}ln\sigma_{k}(0,\mathbf{x_{0}})
\end{equation}

where $k$ is the number of iterations of the map.

As mentioned in prior work~\cite{Tricoche2012b}, identifying a proper iteration scale $k$ is challenging
for the characterization of the structures. However a large enough
number of iterations would ``sharpen'' even the small (visible) structures, the structures
associated with a shorter number of iterations become extremely noisy in the resulting images.

\begin{figure}[!h]
  \centering
  \includegraphics[width=\linewidth]{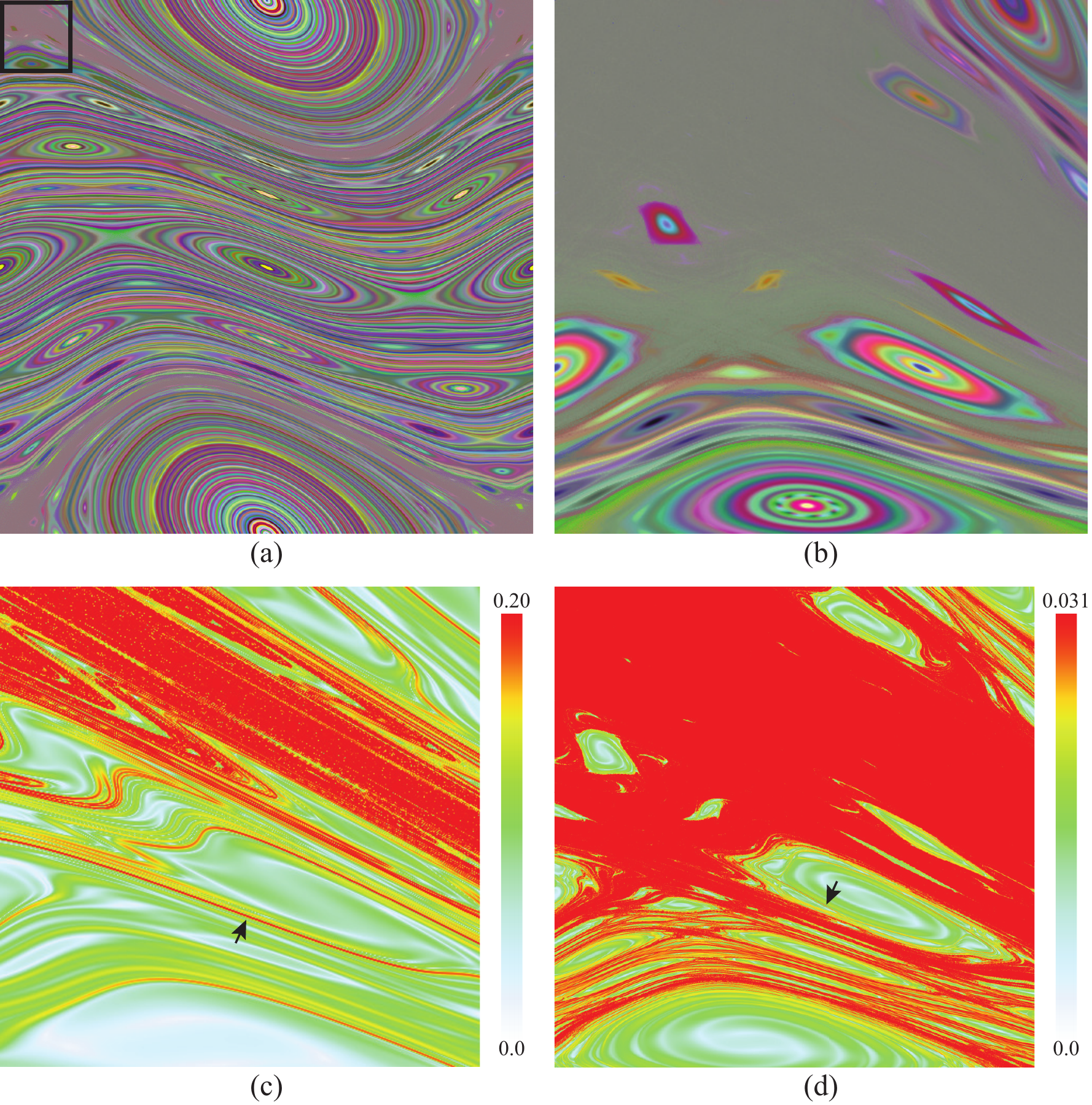}
  \caption{Salient structures of standard map for $K=0.75$. (a) Salient structures is depicted using orbit averaging~\cite{Tricoche2011}. (b) Closed-up in the top-left region highlighted in (a), which near the saddle of period $1$. An island chain associated with this saddle is clearly visualized. (c) FTLE-based visualization of stable manifolds with the number of iterations $k=40$. (d) Stable manifolds at the same region with the number of iterations $k=300$. One may noticed that salient structures (highlighted by the black arrow in (d)) inside isolated islands are visible at such a high number of iterations while others (highlighted by the black arrow in (c)) become hard to distinguish due to the limitation on the spatial sampling resolution.\label{fig:standard_map_ftle}}
\end{figure}

Figure~\ref{fig:standard_map_ftle}(c) and Figure~\ref{fig:standard_map_ftle}(d) visualize
the FTLE measured on standard map for $K=0.75$. Previous study shows that FTLE-based method
is able to convey the chaos that surround the saddle points, and present the invariant
manifolds of the individual island chains of the map~\cite{Tricoche2012b}. However, using
a fixed number of iterations, it is hard to distinguish structures associated with a shorter
number of iterations while keeping the small structures like the ones inside islands visible.

Using the discrete notion of finite-time, our temporal-scale analysis approach provides a
solution in \textit{maps}. It analyzes all FTLE fields computed with different number of
iterations $k$ and characterize salient structures from them. Section~\ref{sec:standard_map_result}
presents the result of our approach.

\section{Results}
\label{sec:6}
Datasets from different applications, including computational fluid dynamics and Hamiltonian systems are considered to study the performance of the presented temporal-scale approach. \textit{Double Gyre}, \textit{Meandering Jet}, and \textit{Standard Map} are well-known synthetic datasets while \textit{Delta Wing} and \textit{Cylinder} are the results of CFD simulations.

\subsection{Double Gyre}
\label{sec:double_gyre_result}
The \textit{Double Gyre} used in this paper is a synthetic test case that we used to document and benchmark the proposed approach. The flow is described by the stream function

\begin{equation}
\psi (x,y,t)=Asin(\pi f(x,t))sin(\pi y)
\end{equation}

where

\begin{equation}
\begin{array}{rcl}
f(x,t) & = & a(t)x^{2}+b(t)x \\
a(t) & = & \varepsilon sin(\omega t) \\
b(t) & = & 1 - 2\varepsilon sin(\omega t)
\end{array}
\end{equation}

over the domain $[0,2]\times [0,1]$. The velocity field is given by

\begin{equation}
\begin{array}{rcccl}
u & = & -\frac{\partial \psi }{\partial y} & = & -\pi A sin(\pi f(x)) cos(\pi y) \\
v & = &  \frac{\partial \psi }{\partial x} & = & \pi A cos(\pi f(x)) sin(\pi y) \frac{df}{dx}
\end{array}
\end{equation}

With $t_{0}=0.0$, $A=0.1$, $\varepsilon=0.25$ and $\omega=\frac{2\pi}{10}$, there is a unstable
manifold (a repelling LCS) attached to the bottom boundary. However as the integration time is
increased, more of the manifold is revealed in the FTLE field. One may notice in
Figure~\ref{fig:standard_ftle_double_gyre} that an early part of the manifold (highlighted in red
rectangle) becomes obfuscated by aliasing and results in a low quality ridge line. The particles released from
this location could reach other structural features in the flow when the integration time is long.
Due to an insufficient spatial sampling resolution, the discontinuity in flow map affects the FTLE
measurement and causes the aliasing in the resulting image. All FTLE fields used in
Figure~\ref{fig:standard_ftle_double_gyre} and Figure~\ref{fig:double_gyre_result} were computed
with a spatial sampling resolution at $2048 \times 1024$.
Comparing Figure~\ref{fig:double_gyre_result}(e) with image (e) and (f) in
Figure~\ref{fig:standard_ftle_double_gyre}, our temporal-scale approach automatically solves the
aliasing problem by using the information in a short integration time and extracts the same high
quality ridge lines at these parts (highlighted in red rectangles in Figure~\ref{fig:double_gyre_result}(e))
 and preserves the manifolds as the ones shown in a long integration time.

\begin{figure}[h]
  \centering
  \includegraphics[width=\linewidth]{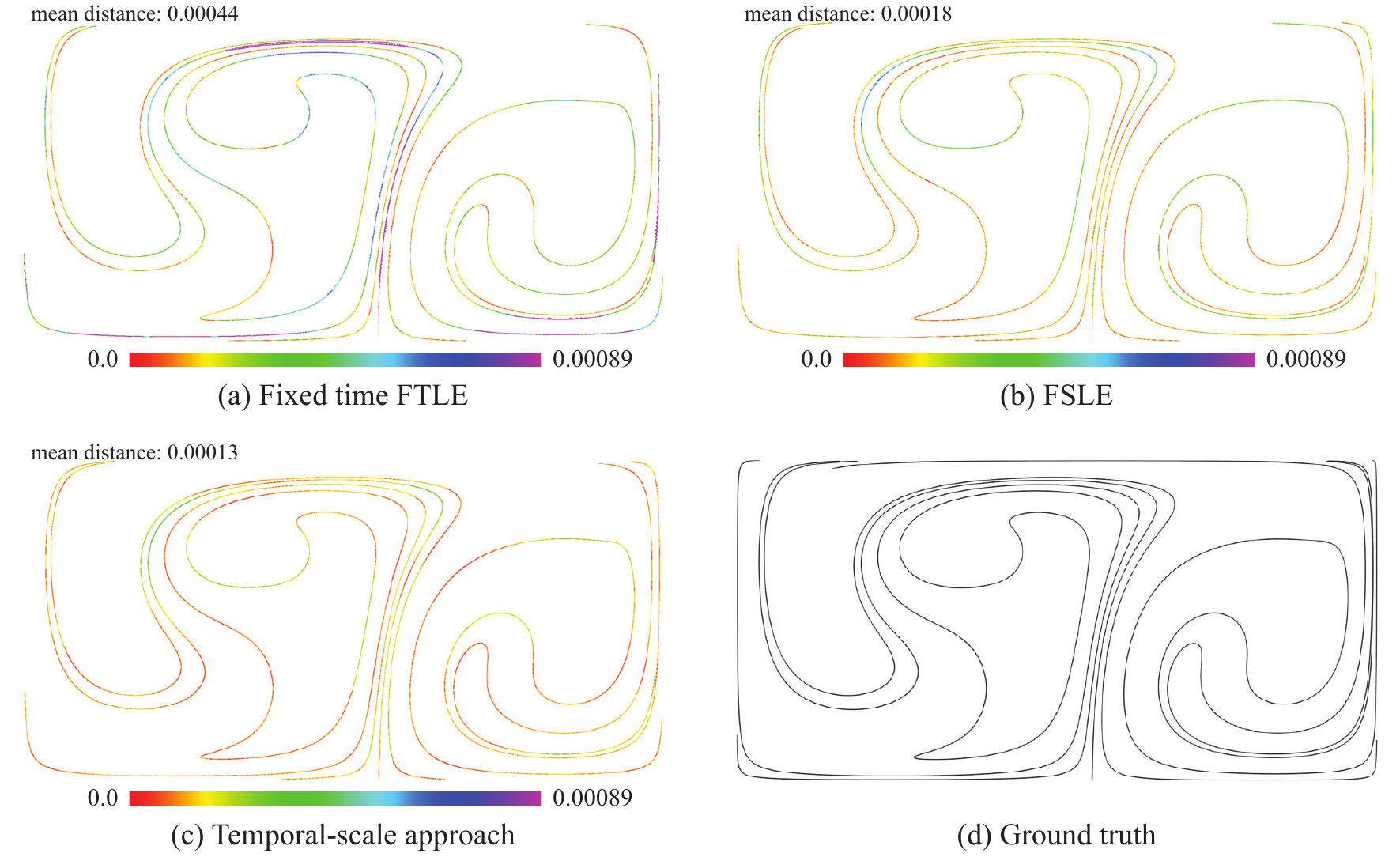}
  \caption{A comparison of LCS characterized by fixed time FTLE, FSLE and the proposed temporal-scale approach. The resulting ridges are color coded by the distance to the ground truth.}
  \label{fig:double_gyre_comparison}
\end{figure}

In order to enable a quantitative comparison between existing methods FTLE-based LCS characterization methods and the proposed temporal-scale approach, the ridge lines extracted from a high spatial sampling
resolution ($16384 \times 8192$) FTLE image with integration time
$\tau=25.0$ were used as the ground truth~\cite{Kuhn2012}. Three methods, namely fixed time FTLE and FSLE, were compared
with the temporal-scale approach in this study. For each method, distances from vertices on
the resulting ridge lines to the ground truth were measured and quantified as the error of
each method. Figure~\ref{fig:double_gyre_comparison} visualizes the error of each method as
well as the ground truth ridge lines. The fixed time FTLE ridges were extracted from the FTLE
field computed with integration time $\tau=25.0$, the FSLE ridges were extracted from the
FSLE field computed with the dispersion factor $300$ and maximum time $400$. The temporal-scale
ridges were extracted from $25$ FTLE fields with a uniform temporal sampling of the integration time
ranging from $1.0$ to $25.0$. All FTLE and FSLE fields were computed with a spatial sampling
resolution at $2048 \times 1024$ and $t_{0}=0.0$.

\begin{figure}[h]
  \centering
  \includegraphics[width=\linewidth]{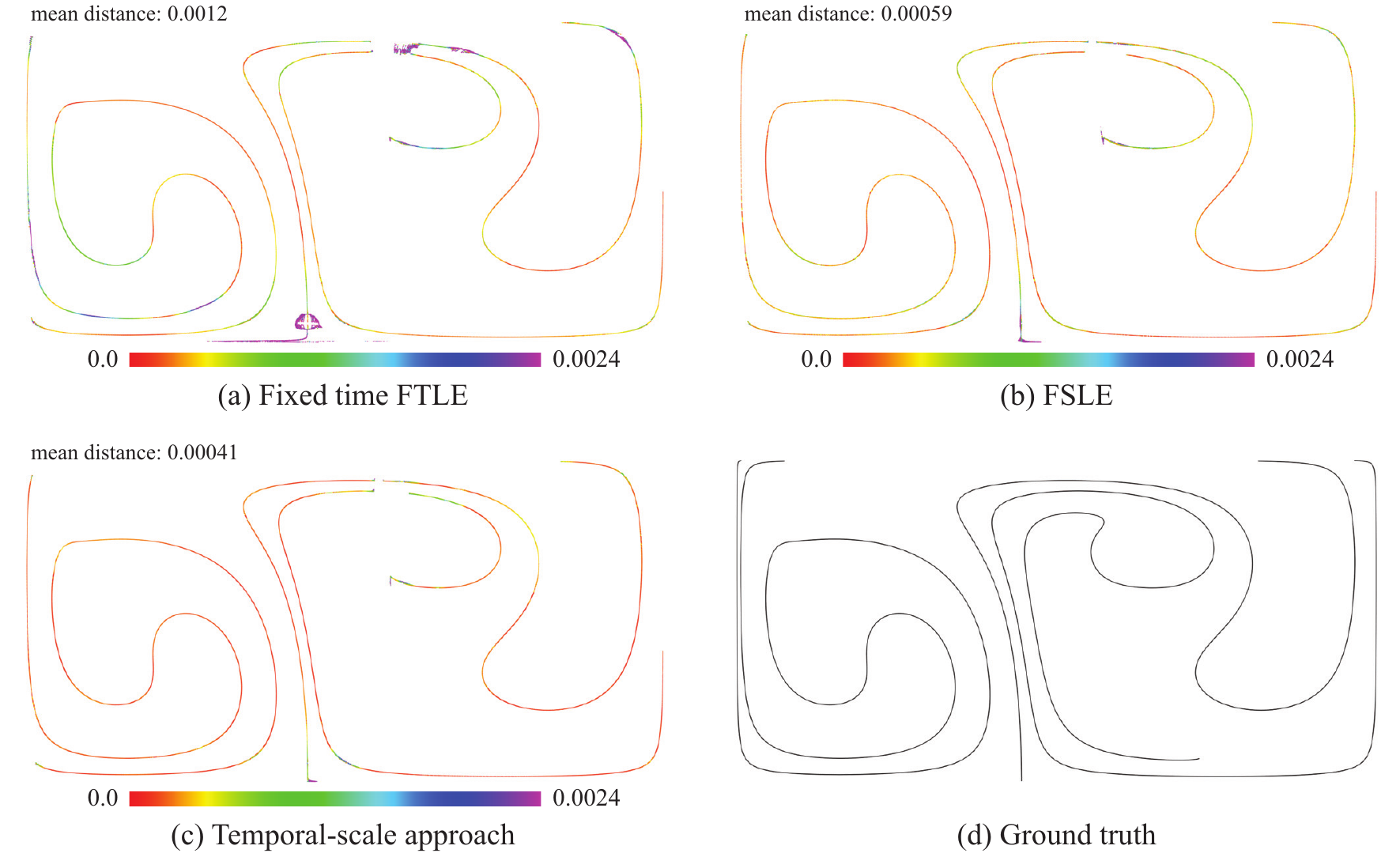}
  \caption{A comparison of the advected LCS at $t=5.0$. All vertices were color coded by the distance to the ground truth.}
  \label{fig:double_gyre_comparison_advected}
\end{figure}

To further study the accuracy of the characterized LCS from these three methods, we advected every
vertices on each initial LCS obtained at $t_{0}=0.0$ to $t=5.0$ and measured
their distance to the ground truth which was achieved in a same way as the previous comparison.
Figure~\ref{fig:double_gyre_comparison_advected} shows the error of the advected LCS initially
obtained from these methods.

With both comparisons, it is clear that temporal-scale approach outperforms both
fixed time FTLE and FSLE under a same spatial sampling resolution. Though the FSLE method could achieve
a close result to the temporal-scale approach, it takes about $2$ hours to compute the FSLE field
on our test machine while the temporal-scale approach only takes a few minutes.

As we already mentioned in section~\ref{sec:hyper_ftle}, the hyper-FTLE definition we purposed which combines the FTLE value and its
associated ridge strength is only one of many possible solutions. Two possible alternatives are considered: FTLE value alone and ridge
strength alone. We extracted ridge lines of both alternatives, and measured the distance to the ground truth.

\begin{table}[h]
\centering
\caption{Mean distance and maximum distance of different hyper-FTLE definition.}
\label{tab:hyper_ftle_definition}
\begin{tabular}{c|c|c}
\hline
 method              & mean distance & maximum distance \\ \hline
 our approach        & $0.00013$     & $0.00072$        \\ \hline
 FTLE value only     & $0.00015$     & $0.0012$         \\ \hline
 ridge strength only & $0.00021$     & $0.0018$         \\ \hline
\end{tabular}
\end{table}

Table~\ref{tab:hyper_ftle_definition} lists both the mean and maximum distance between the extracted ridges with
different hyper-FTLE definitions to the ground truth. Comparing these numbers to the spatial sampling distance, $0.00097$,
only the maximum error of the ridge lines extracted by our approach is less than the spatial sampling distance.

\subsection{Meandering Jet}
\begin{figure}[htb]
  \centering
  \includegraphics[width=\linewidth]{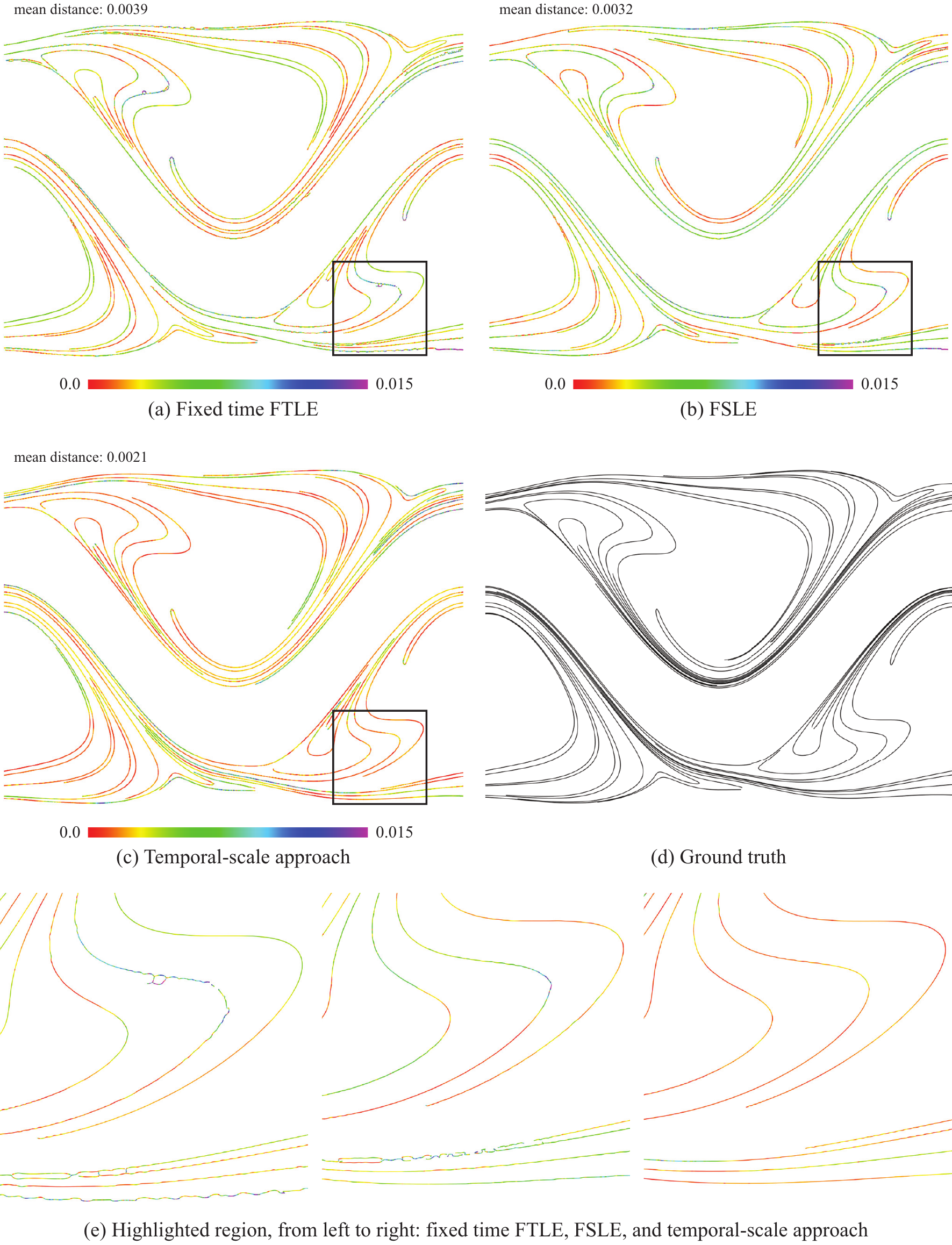}
  \caption{A comparison of LCS characterized by fixed-time FTLE, FSLE and the proposed temporal-scale approach on meandering jet dataset. The resulting ridges are color coded by the distance to the ground truth.}
  \label{fig:meandering_jet_comparison}
\end{figure}

The meandering jet flow has been widely used by FSLE approaches~\cite{Aurell:1997:Predictability}.
A comparison of FTLE and FSLE on this dataset has been studied by Peikert~\cite{Peikert:2014:A-comparison} and Boffetta et al.~\cite{Boffetta:2001:Detecting}.
Both of them used Samelson's model with parameters $B_{0}=1.2$, $L=10.0$, $k=2\pi/L$ and $c=0.1$.
For the parameter pair $(\omega, \varepsilon)$ the setting $(0.1, 0.3)$ was investigated.
We reproduced FSLE results for the same parameter settings with dispersion factor $r=300$, maximum time $300\pi$(fifteen flow periods), extracted ridge lines from FSLE field, fixed time FTLE field with integration time equals to $65\pi$ and compared them with results through the temporal-scale analysis approach (integration time ranging from $30\pi$ to $65\pi$).  All FTLE and FSLE fields were computed with a spatial sampling
resolution at $1000 \times 800$ and $t_{0}=0.0$.
Similar to \textit{Double Gyre}, ridge lines extracted from a high resolution, $4000 \times 3200$, were used as ground truth. Then for each
vertex on the resulting ridge lines, a distance to the ground truth was measured.
Figure~\ref{fig:meandering_jet_comparison} shows the comparison of fixed time FTLE, FSLE, and the temporal-scale approach.

The comparison on meandering jet dataset indicates that the proposed temporal-scale approach has the ability to characterize smoother and more accurate LCS than the existing methods.

\subsection{2D Flow Around a Cylinder}
The 2D flow around a cylinder was simulated by Weinkauf~\cite{weinkauf10c} using the free
software \emph{Gerris Flow Solver}~\cite{gerrisflowsolver}. This third flow dataset constitutes arguably a more challenging test case then the previous two and it allows us to offer additional insight into the benefits of our proposed solution for the characterization of Lagrangian coherent structures.

\begin{figure*}[h]
  \centering
  \includegraphics[width=7.0in]{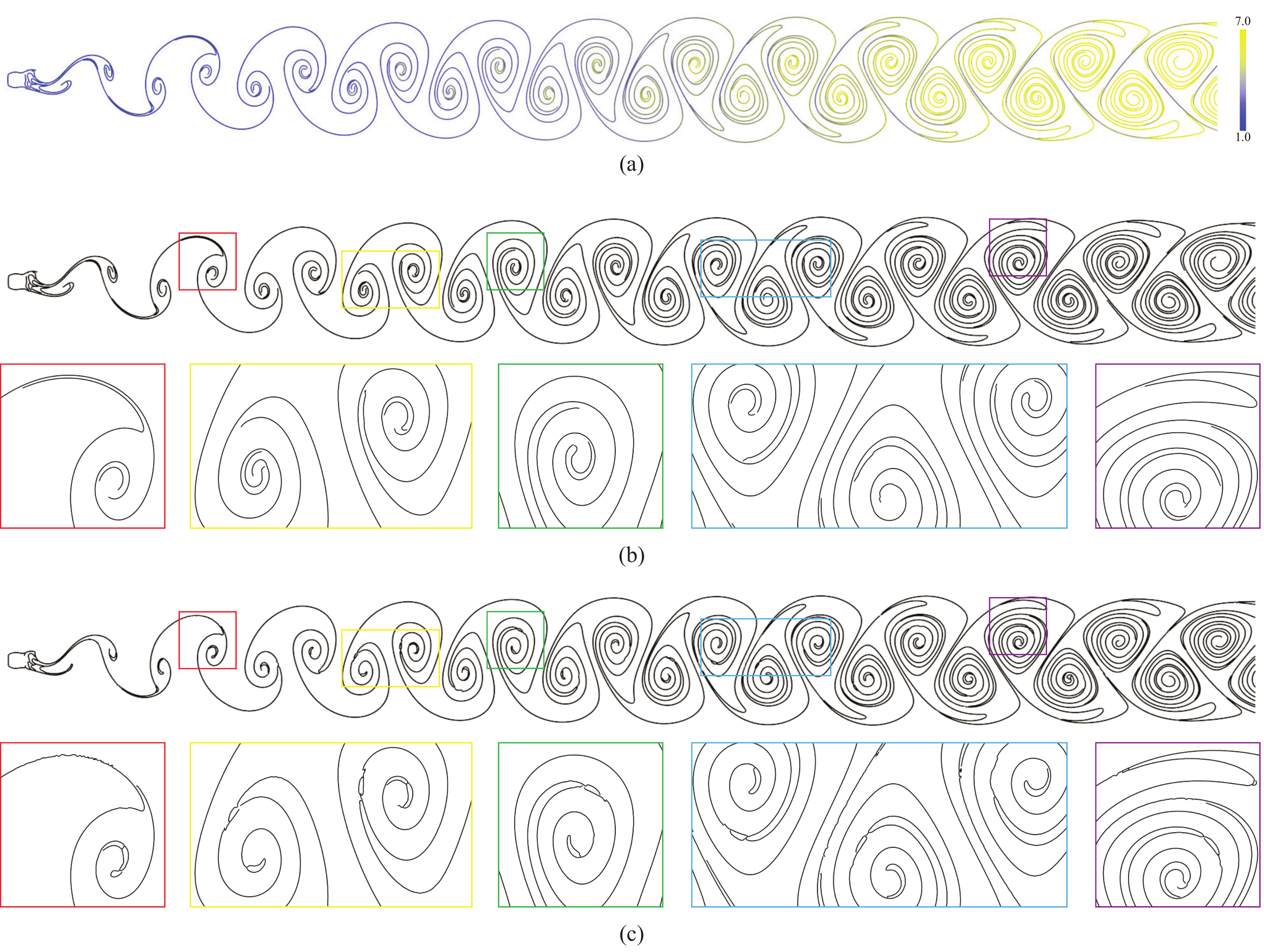}
  \caption{Ridge lines extracted from the 2D cylinder dataset. (a) Ridge lines extracted by our approach, color coded by the integration time ranging from $1.0$ (blue) to $7.0$ (yellow); A detailed comparison of the ridge lines extracted by our approach (b) and
by a standard ridge detection approach on an FTLE field with a fixed integration time $\tau=7.0$ (c) is given.}
  \label{fig:cylinder2D_result_FTLE}
\end{figure*}

Using a large spatial sampling resolution, $7000 \times 1000$, our approach successfully reconstructed LCS
 of the von K\'{a}rm\'{a}n vortex from the 2D simulated data. Figure
\ref{fig:cylinder2D_result_FTLE} shows the ridge lines extracted from the FTLE fields computed
with the integration time ranging from $1.0$ to $7.0$. Image (a) illustrates the LCS formed
at a short integration time (color coded as blue) are embedded with the one formed at a long integration
time (color coded as yellow). A detailed comparison of the ridge lines extracted by our approach and
by a standard ridge detection approach on an FTLE field with a fixed integration time $\tau=7.0$ is given as image (b) and (c) in Figure~\ref{fig:cylinder2D_result_FTLE}.

\begin{figure*}[h]
  \centering
  \includegraphics[width=7.0in]{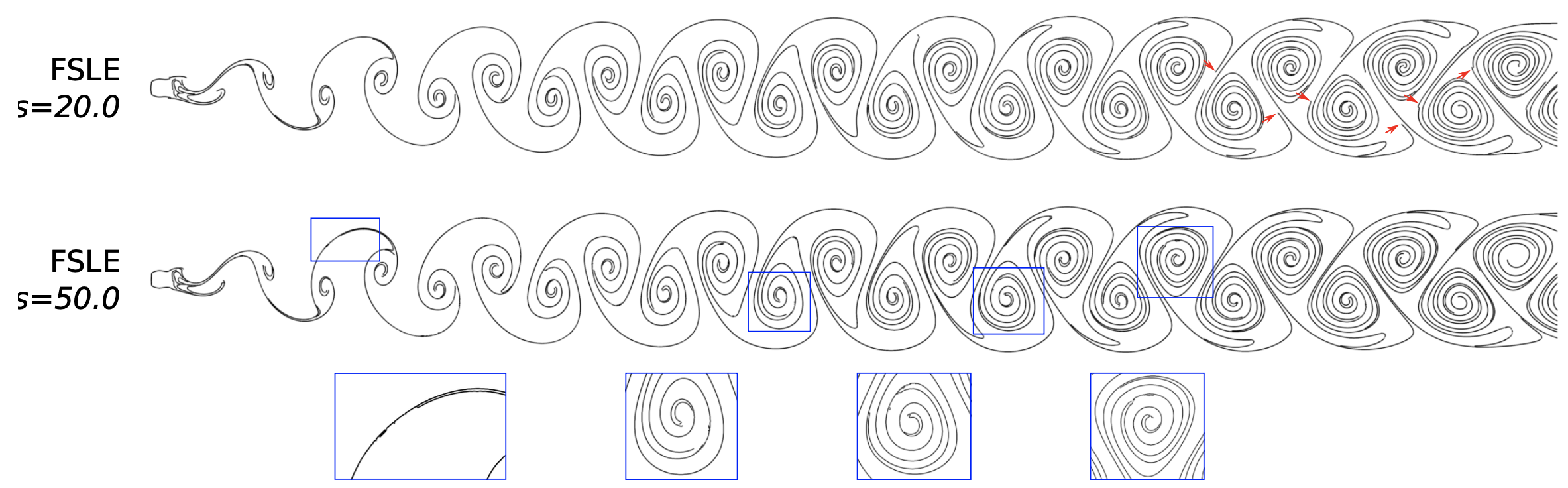}
  \caption{Ridge lines extracted from the FSLE method using two dispersion factors: $\delta=20.0$ and $\delta=50.0$.}
  \label{fig:cylinder2D_result_FSLE}
\end{figure*}

Similar to the FTLE results, the ridges extracted from the FSLE method show that different structures emerge at different values of the dispersion factor. For example, structures highlighted by red arrows in figure~\ref{fig:cylinder2D_result_FSLE} have not emerged until the dispersion factor reaches a large value. Moreover, structures which are revealed at a small dispersion factor become rough and broken at a large dispersion factor (\eg{} structures highlighted by blue rectangles). In addition, Karrasch and Haller have documented some limitations of the FSLE method in Lagrangian coherent structure detection, such as ill-posedness, artificial jump-discontinuities, and sensitivity with respect to the computational time step~\cite{Karrasch:2013:Do-finite-size}.

\begin{figure*}[h]
  \centering
  \includegraphics[width=7.0in]{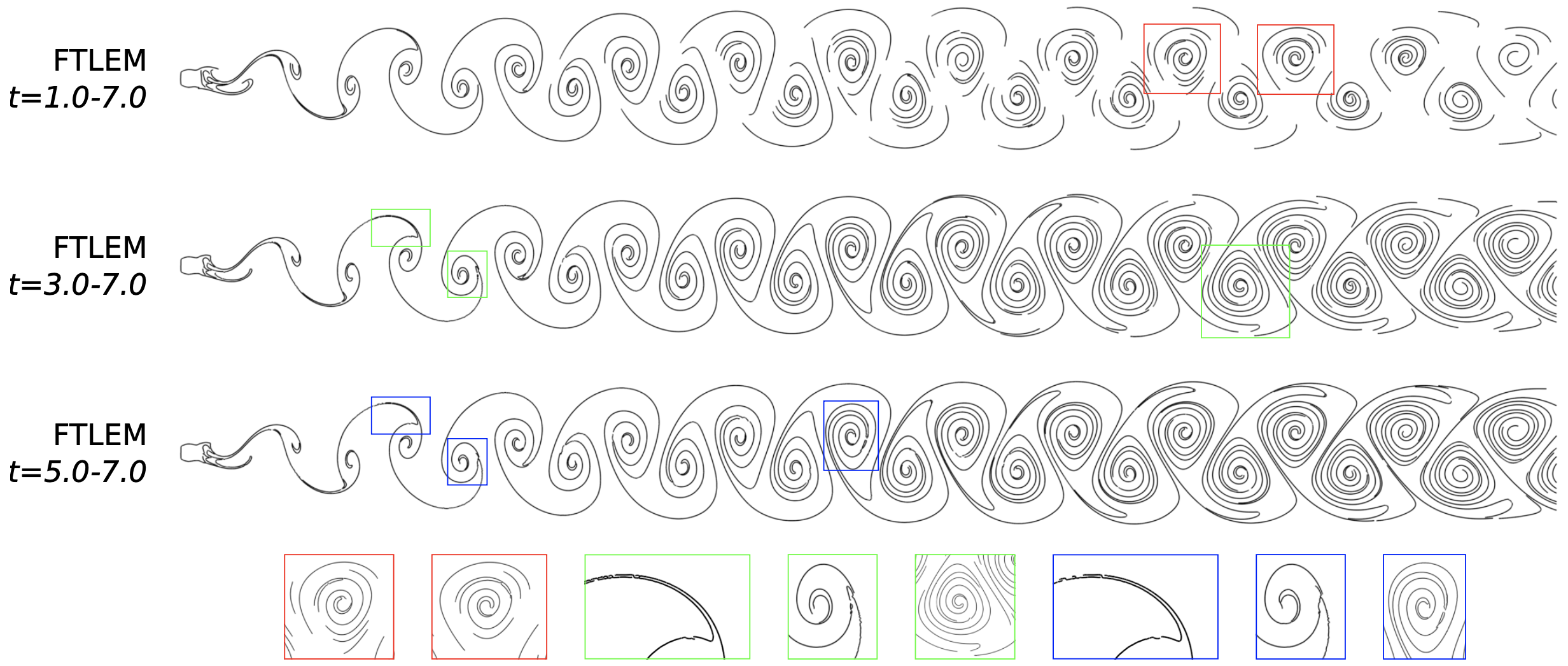}
  \caption{Ridge lines extracted from the FTLEM method using different time intervals: $t \in [1.0, 7.0]$, $t \in [3.0, 7.0]$ and $t \in [5.0, 7.0]$.}
  \label{fig:cylinder2D_result_FTLEM}
\end{figure*}

Finally, comparing our approach with the FTLEM method, one may notice that both methods assign different integration times to different regions of the vector field and capture high expansions along the trajectory instead of only analyzing the final flow map. However, the FTLEM fields visualized in figure 1 show high FTLE values at a short integration time dominate the result of FTLEM method, especially when we apply the method on a large integration time interval (\eg{} $t = 1.0···7.0$ or $t = 3.0···7.0$). The ridge lines extracted from those fields in figure~\ref{fig:cylinder2D_result_FTLEM} show a large number of missing and spurious structures (\eg{} highlighted by red rectangles). One can eliminate such situation by applying the FTLEM method on a small integration time interval (\eg{} $t = 5.0···7.0$). However, structures forming before the lower end of that interval ($t=5.0$) become obfuscated in the results. Indeed, all scalar fields produced by the FTLEM method within that integration time interval prove unable to generate clear and smooth structures (\eg{} highlighted by blue rectangles). In contrast, by considering simultaneously FTLE value and local ridge strength, our method penalizes high FTLE values with low ridge strength values obtained after a short integration time. As a result, our approach enables high-quality structure characterization (the results of our approach are shown in figure~\ref{fig:cylinder2D_result_FTLE}).

\subsection{Delta Wing}
The EDELTA dataset is designed to study the transient flow above a delta wing
at low speeds and increasing angle of attack.
The angle of attack increases over time, leading to vortex breakdown in later timesteps.

\begin{figure}[h]
  \centering
  \includegraphics[width=\linewidth]{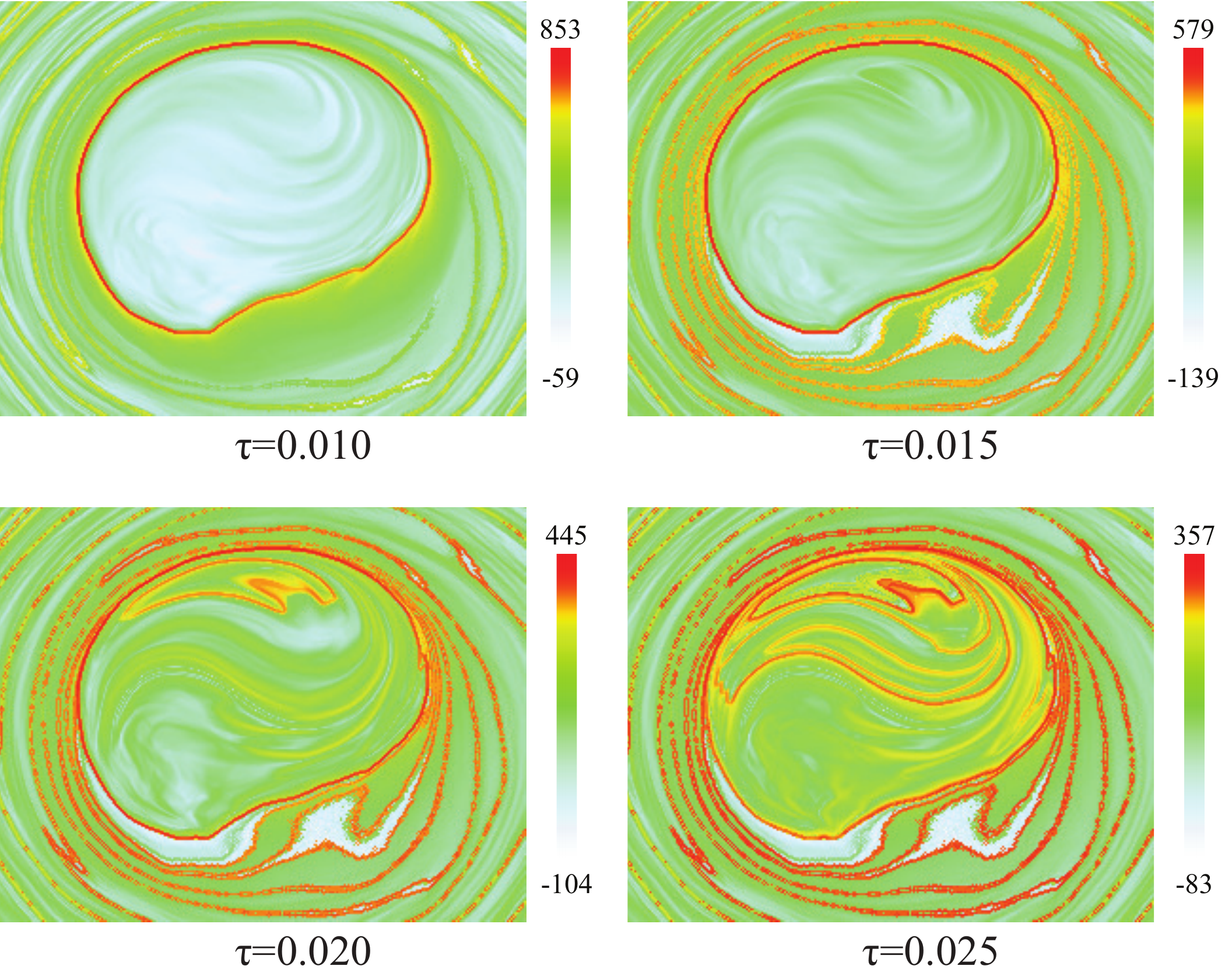}
  \caption{FTLE fields computed with four different integration time.}
  \label{fig:delta_wing_ftle}
\end{figure}

Previous study~\cite{Tricoche:2004:Visualization} has already shown two asymmetric vortex breakdown bubbles exist in this dataset,
a chaotic vortex breakdown is on the left side of the wing while a stable one is on the right side.
The main challenge in characterizing LCS of the vortex breakdown bubble is
the complexity of the flow near the edge of delta wing resulting a noisy FTLE field when the integration time reaches a certain value.
Figure~\ref{fig:delta_wing_ftle} visualizes four slices of backward FTLE fields computed with a different integration time than for the right vortex breakdown bubble.
With the integration time $\tau$ equals to $0.020$ and $0.025$, the noise around the outer layer of the vortex breakdown bubble makes it difficult to extract clear and smooth ridge surfaces. On the other hand, the inner parts of the vortex breakdown can only be revealed at a sufficient long integration time such as $0.025$.

Images (a) and (b) in Figure~\ref{fig:delta_wing_right_result} show the ridge surfaces of the right
vortex breakdown bubble extracted from the FTLE field with a ``short'' integration time
$\tau=0.01$ and a ``long'' integration time $\tau=0.025$ respectively. Image (c) illustrates the results of the time-scale approach with an integration time ranging from $0.005$ to $0.025$. Comparing those images, our approach not only characterizes a high quality ridge surface of the inner part of the vortex breakdown bubble, but also extracts a smooth and complete result for the outer layer. Image (d) visualizes the ridge surfaces rendered with transparency and a different camera setting.



\subsection{Standard Map}
\label{sec:standard_map_result}
The standard map (also known as the Chirikov-Taylor map)~\cite{chirikov1979,chirikov1969}
is an area-preserving chaotic map for two canonical dynamical variables, namely momentum
and coordinate $(p,x)$, from a square with side $2\pi$ onto itself.
The map is defined as follows

\begin{equation}
\begin{matrix}
\overline{p} & = & p+Ksinx \\
\overline{x} & = & x + \overline{p}
\end{matrix}
\end{equation}

\begin{figure}[h]
  \centering
  \includegraphics[width=3.0in]{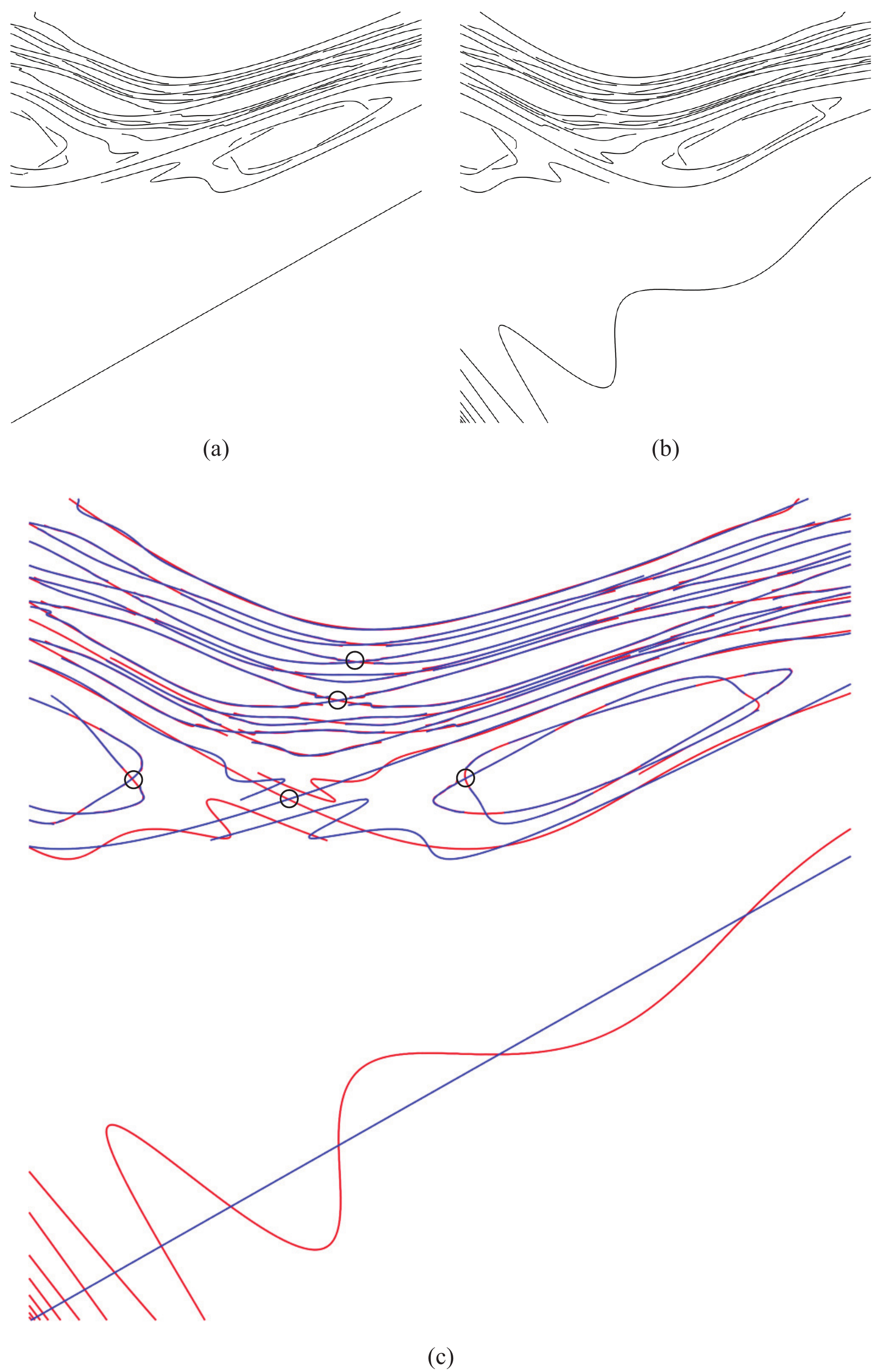}
  \caption{Results of the temporal-scale approach on standard map with $K=0.75$. (a) characterized stable manifolds. (b) characterized unstable manifolds. (c) Fixed points can be detected as the intersection points between stable manifolds (blue) and unstable manifolds (red). }
  \label{fig:standard_map_result}
\end{figure}

Where $p$ and $x$ are taken modulo $2\pi$ and $K$ is the parameter that controls the
nonlinearity of the map. The bar indicates the new values of variables after one iteration.
The standard map describes the dynamics of several mechanical systems and has attracted
the attention of theoretical and computational research alike since it is a simple yet
powerful tool to study Hamiltonian chaos.
We tested the proposed temporal-scale approach on the standard map for $K=0.75$.
Figure~\ref{fig:standard_map_result}(a) and (b) show the characterized stable and
unstable manifolds in the interesting region of Figure~\ref{fig:standard_map_ftle}
by applying the temporal-scale analysis both on the forward and backward time FTLE fields computed
with a uniform sampling of $n$ ranging from $10$ to $300$. In the context of Hamiltonian systems,
finding fixed points in such maps can be a computationally challenging task
that requires an extremely dense sampling of the phase portrait. With highly accurate characterized manifolds,
fixed points can easily be detected by testing the intersection points between stable and unstable manifolds.
Figure~\ref{fig:standard_map_result}(c) illustrates both stable (in blue) and unstable (in red) manifolds
simultaneously in this region. Many intersection points can be identified in this image.
We have tested several intersection points highlighted in black cycles, and computed the distance $\delta d$ between
the coordinate of a intersection point to the actual fixed point. The average distance  of those
highlighted points is around $3\times 10^{-5}$, which is relatively small compared to the sampling distance used in the FTLE computation ($3.8\times 10{-4}$).

\section{Discussion}
\label{sec:7}
\subsection{Performance}
\label{sec:performance}
To study the performance of the proposed temporal-scale approach, we measured the running time on a machine with a hex-core i7 CPU and $12GB$ memory.
Table~\ref{tab:performance} summarizes the running time of the temporal-scale approach on different datasets, as well as the running time of fixed time FTLE and FSLE.
Both the running time of fixed time FTLE and temporal-scale approach include the time used in flow map generation and FTLE measurement.
The temporal sampling resolution means the number of FTLE fields used to sample the integration time within the user defined interval.
Temporal-scale analysis was performed through these FTLE fields.
In unsteady flows, the running time of both fixed time FTLE and temporal-scale approach are close to each other.
It is because the computational time spends on flow map generation dominates the running time in both cases. However, FSLE requires a significant longer
running time as it usually needs a longer integration time to reveal salient structures and has to measure the dispersion during every time steps.

\begin{table*}[h]
\centering
\caption{Measured running time on different datasets.}
\label{tab:performance}
\begin{tabular}{c|c|c|c|c|c}
\hline
 dataset        & spatial sampling resolution & temporal sampling resolution & fixed time FTLE & FSLE      & temporal-scale approach   \\ \hline
 \textit{Double Gyre}    &  $2048 \times 1024$         & $25$                         & $2$ min         & $131$ min & $3$ min   \\ \hline
 \textit{Meandering Jet} &  $1000 \times 800$          & $36$                         & $3$ min         & $84$ min  & $3$ min                 \\ \hline
 \textit{Cylinder 2D}    &  $7000 \times 1000$         & $31$                         & $21$ min        & N/A       & $23$ min                \\ \hline
 \textit{Delta Wing} (left bubble) &  $95 \times 228 \times 189$ & $13$                         & $113$ min       & N/A       & $116$ min               \\ \hline
 \textit{Standard Map}   &  $2048 \times 2048$         & $59$                         & N/A             & N/A       & $4$ min                \\ \hline
\end{tabular}
\end{table*}

\subsection{Temporal resolution}
\label{sec:temporal_resolution}
\begin{figure}[htb]
  \centering
  \includegraphics[width=2.8in]{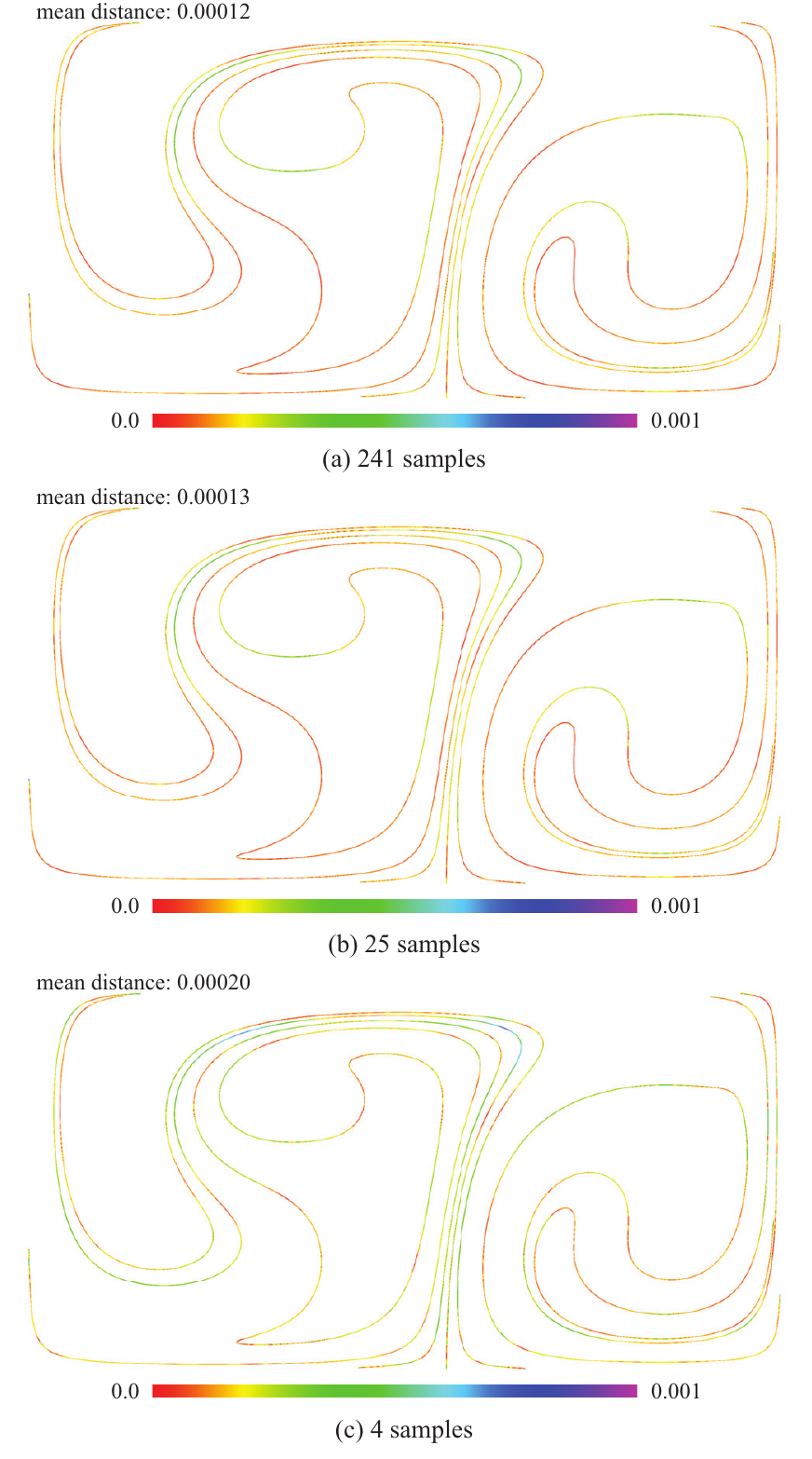}
  \caption{Ridge lines extracted from \textit{Double Gyre} dataset using temporal-scale approach through different number of temporal samples.}
  \label{fig:time_resolution}
\end{figure}

We investigated the influence of the temporal resolution on the result by applying our temporal-scale approach on \textit{Double Gyre} with different temporal resolution settings.
In general, more samples within the user defined integration time interval $[\tau_{LB},\tau_{UB}]$ results in a higher accuracy LCS characterized by our approach. Figure~\ref{fig:time_resolution} visualizes the distance between vertices on extracted ridge lines to the ground truth using different number of samplings within a user defined integration time interval $[1,25]$. The result shows the improvement from $26$ samples to $241$ samples is relatively small compared to the improvement from $4$ to $26$.

\subsection{Applications in Flow Visualization}
\label{sec:integration}
\begin{figure}[htb]
  \centering
   \includegraphics[width=3.5in]{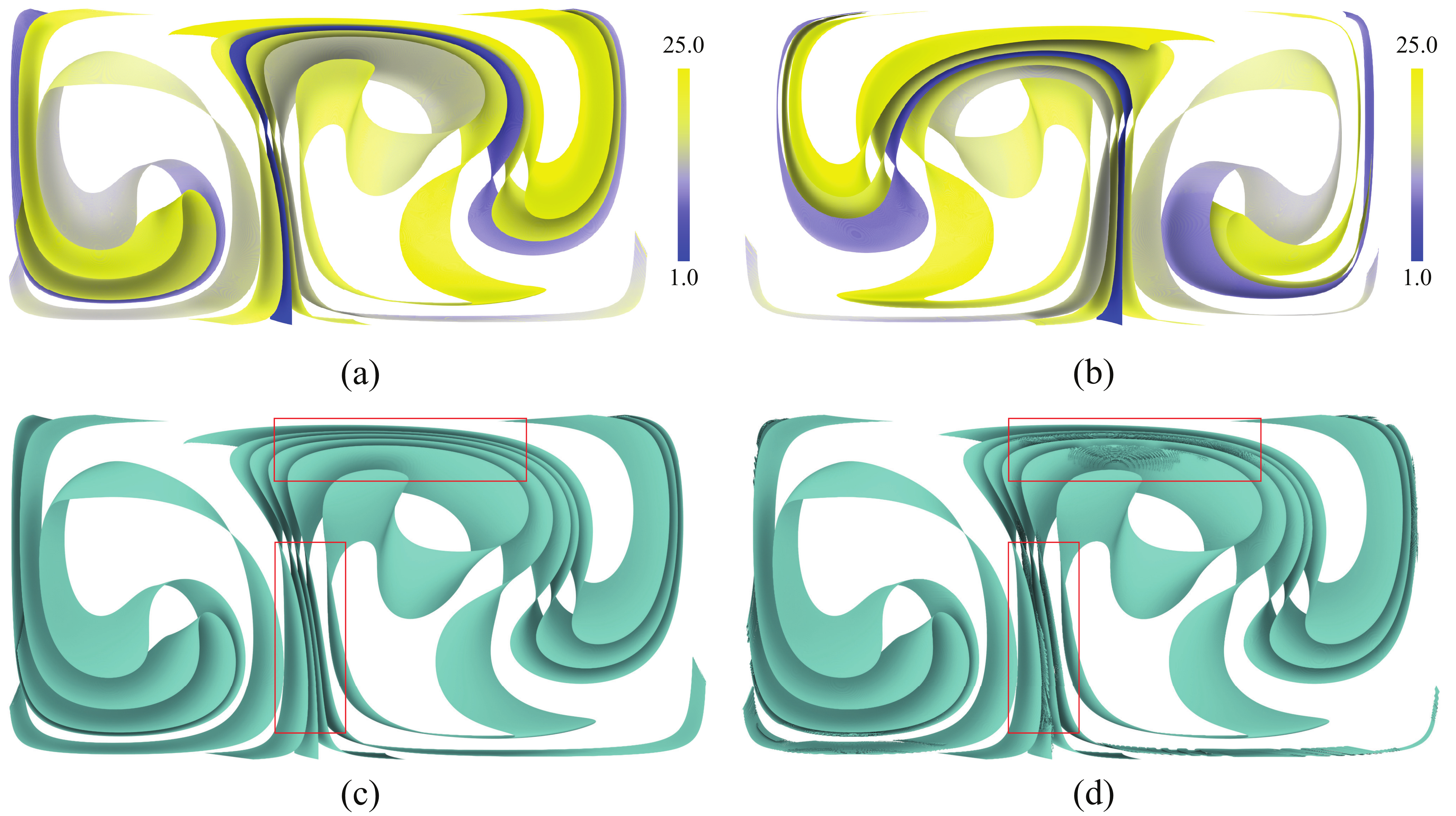}
   \caption{Forward time space-time ridge surfaces characterized by our approach. (a) and (b) shows the rendered space-time ridge surfaces characterized from the \textit{Double Gyre} dataset with integration time ranging from $1.0$ (in blue) to $25.0$ (in yellow) using different camera settings. A detailed comparison of our approach (c) and the space-time ridge surfaces characterized from a FTLE field with a fixed integration time $\tau=25.0$ (d) is given. Differences are highlighted by red rectangles.}
   \label{fig:double_gyre_space_time}
\end{figure}

As the classical FTLE based LCS approach, our temporal-scale approach can easily be integrated into other fluid analysis and visualization methods that are based on LCS. Figure~\ref{fig:double_gyre_space_time} shows the result of our approach in a LCS based space-time visualization framework~\cite{Sadlo2013} for the \textit{Double Gyre} dataset. In this framework, time-dependent vector fields turn into stationary ones by treating time as an additional dimension. Therefore, 2D unsteady vector fields $(u(x,y,t),v(x,y,t))^\intercal$ are converted into a steady 3D vector field $(u(x,y,t), v(x,y,t), t)^\intercal$. Our approach searches the best integration time for each point on the sampling grid and extracts the corresponding ridge surface.

\section{Conclusion}
\label{sec:8}
In many applications of Lagrangian coherent structures, the integration time used to compute FTLE is an important yet ambiguous parameter. In this paper, we have shown that an automatic method to determine the spatially varying optimal integration time enables better LCS characterization results. Instead of showing the structures corresponding to a single and arbitrary integration time, our approach embeds various temporal scales into a time continuum to produce an improved presentation of all LCS within a relevant time range.

The evaluation of the method with both synthetic and real-world datasets shows the ability of our approach to reveal important structural features in time-dependent fluid flows. Further, we have shown the benefits of this approach in the context of maps Hamiltonian systems. In all these cases our method reveals structures that are typically missed by other methods.

An interesting avenue for future work concerns the improvement of the performance of the scale-space approach using a GPU-based implementation and develop a user friendly interface. This would allow the user to interactively visualize and explore the structural features characterized by an approach combining temporal and spatial scales in the analysis of the dataset. The extended time perspective afforded by our method could also find compelling applications in the context of methods that aim to reduce the redundancy of LCS extraction and tracking over time.

\acknowledgments{
This work was supported in part by NSF CAREER Award OCI 1150000 and by a gift by Intel.}

\bibliographystyle{abbrv-doi}
\bibliography{Ziang,xmt}
\end{document}